\documentclass[prd,superscriptaddress,nofootinbib,amsmath,amssymb,aps,10pt]{revtex4}

\usepackage{bm}
\usepackage{amsfonts}
\usepackage{latexsym}
\usepackage[latin1]{inputenc}
\usepackage{graphicx}
\usepackage{amsmath}
\usepackage{palatino}
\usepackage{mathpazo}
\usepackage{epstopdf}

\usepackage{booktabs}
\usepackage{dcolumn}
\usepackage{array}

\usepackage{diagbox}

\usepackage{subfiles} 
\usepackage{ulem} 

\graphicspath{{images/}{../images/}}

\def\jnl@style{\it}
\def\aaref@jnl#1{{\jnl@style#1}}

\def\aaref@jnl#1{{\jnl@style#1}}

\def\aj{\aaref@jnl{AJ}}                   
\def\apj{\aaref@jnl{ApJ}}                 
\def\apjl{\aaref@jnl{ApJ}}                
\def\apjs{\aaref@jnl{ApJS}}               
\def\apss{\aaref@jnl{Ap\&SS}}             
\def\aap{\aaref@jnl{A\&A}}                
\def\aapr{\aaref@jnl{A\&A~Rev.}}          
\def\aaps{\aaref@jnl{A\&AS}}              
\def\mnras{\aaref@jnl{Mon.~Not.~Roy.~Astron.~Soc.}}             
\def\prd{\aaref@jnl{Phys.~Rev.~D}}        
\def\prc{\aaref@jnl{Phys.~Rev.~C}}  
\def\prl{\aaref@jnl{Phys.~Rev.~Lett.}}    
\def\qjras{\aaref@jnl{QJRAS}}             
\def\skytel{\aaref@jnl{S\&T}}             
\def\ssr{\aaref@jnl{Space~Sci.~Rev.}}     
\def\zap{\aaref@jnl{ZAp}}                 
\def\nat{\aaref@jnl{Nature}}              
\def\aplett{\aaref@jnl{Astrophys.~Lett.}} 
\def\apspr{\aaref@jnl{Astrophys.~Space~Phys.~Res.}} 
\def\physrep{\aaref@jnl{Phys.~Rep.}}      
\def\physscr{\aaref@jnl{Phys.~Scr}}       
\def\commat{\aaref@jnl{Comm.~Math.~Phys.}}              
\def\science{\aaref@jnl{Science}}               
\def\cqg{\aaref@jnl{Classical Quant.~Grav.}}            
\def\jpcs{\aaref@jnl{JPCS}}                                     
\def\ijmpd{\aaref@jnl{Int.~J.~Mod.~Phys.~D}}                    
\def\grg{\aaref@jnl{Gen.~Relat.~Gravit.}}               
\def\rpp{\aaref@jnl{Rep.~Prog.~Phys.}}          
\def\npa{\aaref@jnl{Nucl.~Phys.~A}}        
\def\lrr{\aaref@jnl{Living Rev.~Rel.}}                   
\def\jcap{\aaref@jnl{J.~Cosmology Astropart.~Phys.}}    
\def\rmp{\aaref@jnl{Rev.~Mod.~Phys.}}   
\def\rmp{\araa@jnl{Annu.~Rev.~Astron.~Astrophys.}}   

\allowdisplaybreaks[1]

\newcommand{\GetGRls}{continues line} 
\newcommand{\GetGRc}{orange} 
\newcommand{\GetSTTmZero}{red} 
\newcommand{\GetSTTmMtwo}{blue} 
\newcommand{\GetSTTmMthree}{green} 
\newcommand{\GetSTTlambdaZero}{dashed line} 
\newcommand{\GetSTTlambdaMMone}{dashed line with one dot} 
\newcommand{\GetSTTlambdaMzero}{dashed line with two dots} 
\newcommand{\GetSTTlambdaMPone}{dashed line with three dots} 

\begin{document}
    
    \title{Moment of inertia -- mass universal relations for neutron stars in scalar-tensor theory with self-interacting massive scalar field}
    
    \author{Dimitar Popchev}
    \affiliation{Department of Theoretical Physics, Faculty of Physics, Sofia University, Sofia 1164, Bulgaria}
    
    \author{Kalin V. Staykov}
    \email{kstaykov@phys.uni-sofia.bg}
    \affiliation{Department of Theoretical Physics, Faculty of Physics, Sofia University, Sofia 1164, Bulgaria}
    
    \author{Daniela D. Doneva}
    \email{daniela.doneva@uni-tuebingen.de}
    \affiliation{Theoretical Astrophysics, Eberhard Karls University of T\"ubingen, T\"ubingen 72076, Germany}
    \affiliation{INRNE -- Bulgarian Academy of Sciences, 1784  Sofia, Bulgaria}
    
    \author{Stoytcho S. Yazadjiev}
    \email{yazad@phys.uni-sofia.bg}
    \affiliation{Department of Theoretical Physics, Faculty of Physics, Sofia University, Sofia 1164, Bulgaria}
    \affiliation{Institute of Mathematics and Informatics, Bulgarian Academy of Sciences, Acad. G. Bonchev Street 8, Sofia 1113, Bulgaria}

    
    \begin{abstract}
        We are investigating universal relations between different normalisations of the moment of inertia and the compactness of neutron stars in slow rotation approximation. We study the relations in particular class of massive scalar-sensor theories with self-interaction, for which significant deviations from General Relativity are allowed for values of the parameters that are in agreement with the observations. Moment of inertia-compactness relations are examined for different normalisation of the moment of inertia. It is shown that for all studied cases the deviations from EOS universality are small for the examined equations of state. On the other hand the scalarization can lead to large deviations from the general relativistic universal relations for values of the parameters that are in agreement with the current observations that can be potentially used to set further test the scalar-tensor theories.
    \end{abstract}
    
    \maketitle
    \date{}
    
    \section{Introduction}
    
    \paragraph*{} The recent direct observations of gravitational waves mark the beginning of new branch in astronomy -- gravitational wave astronomy \cite{Abbott:2016blz,Abbott:2016nmj,Abbott:2017gyy,Abbott:2017vtc,Abbott:2017oio,GWdetection}. With the upcoming start of the next generation gravitational wave detectors and radio astronomy observatories, which will greatly extend the astrophysical phenomena that can be detected, the prospects in front of astrophysics look ever so promising. On another front, one of the most important challenges modern astrophysics faces today is to provide answer what causes the observed accelerated expansion of the Universe. The attempts in solving this problem can be summarized into two main ways. One is the introduction of a new type of matter with exotic properties, which manifests itself only via gravitational interaction with visible matter. The other alternative way is to construct generalized gravitational field theory, which admits GR as weak field approximation on astrophysical scales, but manifests itself in qualitatively different way on cosmological scales and possibly in the strong field regime. 
    
    \paragraph*{} The area of research of alternative theories of gravity is very active and many different attempts to generalize the Einstein's theory have been made, but only small portion of them can pass all observational tests. One such class of theories presents an natural cosmological and astrophysical generalization of GR -- scalar-tensor theories of gravity (STT) \cite{Fierz56, Jordan59, brans611, Damour1992, Will1993}, in which a scalar field is included as an additional mediator of the gravitational interaction, apart from the spacetime metric. The presented subclasses of STT, with an Einstein frame coupling function of the form  $ \alpha(\varphi) = \beta \varphi $, are of particular interest, due the fact that they are indistinguishable from GR in the weak field regime, but show nonperturbative effect -- spontaneous scalarization \cite{Damour1993, Damour1996, Harada1997, Harada1998,Salgado1998, Pani2011, Sotani2012, Doneva2013, Motahar2017, Popchev2015,Berti2015b,Doneva2017}, in which large deviations from GR are observed, in the strong field regime, e.g. in the gravitational field of compact objects like neutron stars (NS). The structure, properties, and physical effects of NSs in such classes of STT were extensively studied in the past decades (see e.g. \cite{Damour1993,Damour1996,Harada1997,Harada1998,Salgado1998,Pani2011,Sotani2012,Doneva2013,Motahar2017}) both in the static and rapidly rotating cases. A few years ago particular interest attracted the STT with massive scalar field \cite{Popchev2015,Ramazanoglu2016,Yazadjiev2016,Doneva:2016xmf} due to the possibilities for much larger deviations from GR compared to the massless case within the observationally allowed values of the parameters.
    
    Current observations of binary systems constrain significantly the parameter space of massless STT  to $ \beta \gtrsim -4.5$ \cite{Demorest10,Freire2012,Antoniadis2013}. For such values of $ \beta $ no significant  deviation from pure GR is observed as far as static neutron stars are considered. The situation changes if we extend the study to STT with a massive scalar field since for a mass $ m_\varphi $ one can assign a Compton wave-length $ \lambda_\varphi = 2\pi/m_{\varphi} $ beyond which the scalar field is exponentially suppressed. For this reason, as shown in \cite{Yazadjiev2016}, the observationally allowed values for $ \beta $ can significantly differ from the massless STT and NS can have considerably different properties and structure (see e.g. \cite{Popchev2015,Ramazanoglu2016,Yazadjiev2016,Doneva:2016xmf,Motahar2017}). The inclusion of an additional quartic self-interaction term in the potential suppresses further the scalarization effect and thus can reconcile even a wider range of values of the parameters with the observations. STT with a self-interacting massive scalar field is specified by three free parameters -- the coupling coefficient $ \beta $, the scalar field mass $ m_\varphi $ and the self-interaction coefficient $ \lambda $. In certain observationally allowed ranges of the parameters $(\beta, m_{\varphi}, \lambda) $ the NS models can have considerably distinct structure compared to GR \cite{StaykovPopchev2018}.
    
    \paragraph*{} A large number of alternative theories of gravity lead either to negligible effects on the NS properties, or the effects fall well within  the  equations of the state (EOS) uncertainty for the matter in the star, that is a drawback in the attempts to constrain the strong field regime of gravity. The large uncertainty in the EOS on one hand is due to our poor understanding of the fundamental interactions that take place at the extreme densities found in the core of NS, and on the other hand -- the lack of enough accurate observational data or the lack of unique interpretation of these data. In order to successfully constrain EOS, observations of the neutron star mass, radius, moment of inertia or tidal Love numbers are normally employed, but currently this is done either with limited accuracy or only a few astrophysical objects with the desired properties are observed (see e.g. \cite{Lattimer2005,Lattimer:2013hma,Ozel:2016oaf}). Since binary NSs are one of the most promising sources of gravitational waves, it is expected that the constraints on the EOS will be quickly improved due to the rapid advance in the gravitational wave astronomy \cite{Abbott:2018exr,Bauswein:2017vtn,Annala:2017llu,Rezzolla:2017aly,Ruiz:2017due}.
    
    \paragraph*{} Workaround for the EOS uncertainty is to use independent, from the EOS, relations of stellar parameters, the so called universal relations. One of the first construction of such relations which involve the NS oscillation frequencies on one hand and the neutron star mass and radius on the other, was made in \cite{Andersson96,Andersson98c,Andersson98a}. Later, these studies have been extended  with inclusion of additional realistic EOS \cite{Benhar04}. In a recent work Lau et. al. \cite{Lau2010} exchanged the compactness with an ``effective compactness'' $ \eta \equiv \sqrt{M^3/I} $ with the purpose of achieving better EOS independence. In \cite{Chirenti2015} one can also find nice investigation of the above mentioned relations and comments on their universality. Another promising universal relations was found by Yagi and Yunes \cite{Yagi2013,Yagi2013a} which connects the normalized moment of inertia $ \bar I $, the tidal Love numbers, and the normalized quadrupole moment $ \bar Q $. This relation was thoroughly investigated in following years in many papers (see e.g. \cite{Maselli2013, Doneva2014, Pappas2014, Chakrabarti2014, Haskell2014, Urbanec2013, Majumder2015, Yagi2015}).
    
    \paragraph*{} The study of different universal relations in alternative theories of gravity has proven to be very fruitful (see e.g. \cite{Sotani04,Sotani2005,Yagi2013a, Sham2014, Kleihaus2014, Pani2014, Doneva2015}).  In some cases clear distinction with GR can be observed that serves as a way to constrain the strong field regime of gravity in an EOS independent way while in other cases the results are not only to some degree EOS independent, but up to a large extend theory independent too. In the latter case one can use the universal relations to determine the NS parameters without any arbitrariness.
    
    \paragraph*{} A  particularly interesting application of the universal relations between normalized moment of inertia and the stellar compactness was employed by Lattimer and Schutz \cite{Lattimer2005}, where they point out the possibility to estimate the NS radius by using the mass and the moment of inertia of a pulsar in binary system. In recent paper Breu and Rezzolla \cite{Breu2016} thoroughly investigate such relations and comment on their application. Following their work Staykov et al \cite{Staykov2016} extended the study by looking at universal relations between normalized moment of inertia and stellar compactness in GR, $(R)$ and STT theories of gravity. 
    
    \paragraph*{} The paper is organized as follows. In Section  II we review the basic theoretical background behind the calculation of neutron star models in STT with massive self-interacting scalar field. In Section III the numerical results are presented. The paper ends with Conclusions.
    \section{Analytical basis}
    
    \paragraph*{} We will adopt the well established in the literature way of examining STT theories by writing out the field equations here in the more convenient non-physical Einstein frame, which is specified with metric $ g_{\mu\nu} $, but we will present the results in following sections in the physical Jordan frame, which is specified with metric $ ^*g_{\mu\nu} $. The two metrics are connected with conformal transformation $ ^*g_{\mu\nu} = \mathcal{A}^{2}(\varphi) g_{\mu\nu}$, were the Einstein frame scalar field is denoted by $ \varphi $. As a rule we will use superscript star $ ^* $ to note quantities in Jordan frame. The general form of the action of STT in the Einstein frame is given as
    \begin{eqnarray}
    \label{Action_STT} S =
    \frac{1}{16\pi G} \int d^4x \sqrt{-g}\left[ R - 2
    g^{\mu\nu}\partial_{\mu}\varphi \partial_{\nu}\varphi - V(\varphi)
    \right] 
    + S_{\rm matter}(\mathcal{A}^2(\varphi)g_{\mu\nu},\chi),
    \end{eqnarray}
    where $ R $ is the Ricci scalar curvature with respect to $ g_{\mu\nu} $. The STT is fully specified by the functions $ \mathcal{A}(\varphi) $ and $ V(\varphi) $. In the present paper we will restrict our study to STT with conformal factor of the form
    \begin{eqnarray}
    \label{CouplingF_STT} \mathcal{A}(\varphi) = e^{\frac{1}{2}\beta \varphi^2},
    \end{eqnarray}
    where $ \beta $ is a parameter. This class of STT is indistinguishable from pure GR in the weak field regime, while exhibiting nonperturbative effects for strong fields. We will adopt a natural non-negative scalar potential with self-interaction
    \begin{eqnarray}
    \label{VF_STT} V(\varphi) = 2m_\varphi^{2} \varphi^{2} + \lambda \varphi^{4},
    \end{eqnarray}
    where $ m_{\varphi} $ is the mass of the scalar field $ \varphi $ and $ \lambda \geq 0 $ is the self-interaction parameter with dimensions $ length^{-2} $.
    
    \paragraph*{} The field equations that follow from (\ref{Action_STT}) are
    \begin{eqnarray}
    \label{FieldR_STT} &&R_{\mu\nu} - \dfrac{1}{2}g_{\mu\nu}R =
    8\pi G T_{\mu\nu} + 2 \nabla_{\mu}\varphi\nabla_{\nu}\varphi - g_{\mu\nu}g^{\alpha\beta}\nabla_{\alpha}\varphi\nabla_{\beta}\varphi - \dfrac{1}{2}V(\varphi)g_{\mu\nu},
    \\
    \label{Fieldvarphi_STT} &&\nabla_{\mu}\nabla^{\mu} \varphi =
    -4\pi G \alpha(\varphi)T + \dfrac{1}{4}\dfrac{dV(\varphi)}{d\varphi},
    \end{eqnarray}
    where $ \nabla_{\mu} $ is the covariant derivative with respect to $ g_{\mu\nu} $, the coupling function $ \alpha(\varphi) $ is defined via the conformal factor $ \alpha(\varphi) = \frac{d\ln A(\varphi)}{d\varphi} $, and $ T $ is the trace of the energy-momentum tensor. Using the field equations and the contracted Bianchi identities we derive the conservation law of the energy-momentum tensor in the Einstein frame:
    \begin{eqnarray}
    \label{TconservE_STT} \nabla_{\mu}T^{\mu}_{\phantom{a} \nu} = \alpha(\varphi) T \nabla_{\nu}\varphi.
    \end{eqnarray}
    
    \paragraph*{} The Einstein frame energy-momentum tensor $ T_{\mu\nu} $ and the Jordan frame one $ ^{*}T_{\mu\nu} $ are related in the following way $ T_{\mu\nu} = A^{2}(\varphi)^{*}T_{\mu\nu} $. We will consider stationary and axisymmetric perfect fluid and scalar field configurations, for which the transformation of the  energy density $ \rho $, the pressure $ p $ and the 4-velocity $ u_{\mu} $ between the Einstein and the Jordan frame are given as follows:
    \begin{eqnarray}
    \label{Other_Rel} \rho = A^{4}(\varphi) ^{*}\rho,\qquad p = A^{4}(\varphi)^{*}p,\qquad u_{\mu} = A^{-1}(\varphi)^{*}u_{\mu}.
    \end{eqnarray}
    
    \paragraph*{} Our metric ansatz is the standard one for stationary and axisymmetric spacetime in the slow rotation approximation  \cite{Hartle1967}:
    \begin{eqnarray}
    \label{MetricAnsatz} ds^{2} = -e^{2\Phi(r)}dt^{2} + e^{2\Lambda(r)}dr^{2} + r^{2}(d\theta^{2} + \sin^{2}\theta d\vartheta^{2}) - 2\omega(r,\theta)r^{2}\sin^{2}\theta d\vartheta dt.
    \end{eqnarray}
    Since the rotational corrections to the metric functions (except for $ \omega(r,\theta) $), the scalar field, the fluid energy density and the pressure are of order $ \mathcal{O}(\Omega^{2}) $, where $ \Omega = u^{\vartheta}/u^{t} $ is the angular velocity, this approximation allows us to calculate the moment of inertia of the star, while the rest parameters, such as the mass and the radius, will coincide with the static case.
    
    \paragraph*{} To obtain the results in the present paper, we solve numerically the dimensionally reduced ODE system derived from eqs. (\ref{FieldR_STT}) and (\ref{Fieldvarphi_STT}) with the metric ansatz (\ref{MetricAnsatz}) by containing at most terms linear in $ \Omega $ and supplementing it with the equation for hydrostatic equilibrium and an EOS. Further details on the mathematical formulation of the problem one can find in \cite{Yazadjiev2014, Staykov2014, StaykovPopchev2018}.
    
    \paragraph*{} In the next section, where we present our numerical results, we shall use dimensionless parameters $ m_{\varphi} \rightarrow m_{\varphi}R_{0} $ and $ \lambda \rightarrow \lambda R_{0}^{2} $, where $ R_{0} = 1.47664 \mbox{km}$ is one half of the solar gravitational radius.
    
    \section{Numerical results}
    
    \subsection{Preliminaries}
    
    \paragraph*{} As discussed earlier, STT with self-interacting massive scalar field with conformal factor $ \mathcal{A}(\varphi) = e^{\beta\varphi^{2}/2} $ has three free parameters -- $ (\beta, m_{\varphi}, \lambda) $ and for different sets of values, it has varying degrees of development of spontaneous scalarization. Namely, decreasing the value of $ \beta < 0 $ increases the deviations from GR, while the increase of either $ m_{\varphi} $ or $ \lambda $ suppresses them \cite{StaykovPopchev2018}. The constraints on the parameter space are derived after confronting against the observations of the gradual orbit contraction of binary pulsars due to the gravitational wave emission and the strongest constraints come from \cite{Freire2012, Antoniadis2013}. The present data on the rate of orbital decay matches very well the GR predictions, which suggests non or negligible scalar gravitational radiation, and thus non or very weak scalarization effect. As result, for the massless STT, $ (\beta, m_{\varphi} = 0, \lambda = 0) $, the observationally allowed values of $ \beta > -4.5 $ are such that possibility only for small deviations from GR is left, since spontaneous scalarization occurs roughly for $ \beta < -4.35 $ \cite{Damour1996, Harada1998} and $ \beta < -3.9 $ \cite{Doneva2013} for the static and the rapidly rotating cases correspondingly.
    
    \paragraph*{} If, however, one considers STT with massive scalar field, $ (\beta, m_{\varphi}, \lambda = 0) $, the parameter space that is in agreement with the same binary observations is substantially expanded. The reason is that the mass of the scalar field suppresses the emission of scalar radiation, which reconciles already discarded values of $ \beta $. The lower boundary of the scalar field mass can be estimated using the distance between the two companions $ r_{b} $ as follows: a negligible scalar gravitational radiation implies that the Compton wavelength should be much smaller than the orbital separation $ \lambda_{\varphi} \ll r_{b}$, which for the observed binaries $ r_{b} \sim 10^{9}\mbox{m}$ translates into $ m_\varphi \gg 10^{-16}\mbox{eV} $. The upper limit is calculated by the condition that the mass of the scalar filed should be such that it does not suppress the spontaneous scalarization in the stars, i.e. the characteristic length of the star should be smaller than the Compton wavelength, which leads to $ m_\varphi \lesssim 10^{-9}\mbox{eV} $. Thus, we will work with the following range of values for $ m_{\varphi} $:
    \begin{eqnarray}
    \label{STT_m_interval} 10^{-16}\mbox{eV} \lesssim m_{\varphi} \lesssim 10^{-9} \mbox{eV}
    \end{eqnarray}
    or in dimensionless units $ 10^{-6} \lesssim m_{\varphi} \lesssim 10 $. Although, there are additional midrange constraints for the mass of the scalar field, the above ones are most reliable and we will stick to them. For such scalar field masses the observationally allowed ranges of values for $ \beta $ significantly increases compared to the massless case, more precisely $ 3 \lesssim -\beta \lesssim 10^{3} $ coming from the requirement that we can have scalarized NS, but no scalarization for white dwarfs. 
    
    \paragraph*{} The allowed range of parameters extends even further, as shown in  \cite{StaykovPopchev2018}, if we consider STT with self-interacting massive scalar field, $ (\beta, m_{\varphi}, \lambda) $. In such STT the scalarization is suppressed further by the self-interaction term and up to a large extent the self-interaction constant $ \lambda $ in (\ref{VF_STT}) has qualitatively very similar effect on the NS properties as the scalar field mass. The main difference comes from the fact that the self-interaction does not change (for fixed mass of the scalar field) the critical values of the parameters where new branches of scalarized solutions originate from the GR ones, while the mass of the scalar field changes these bifurcation points.
    
    \paragraph*{} In this paper we will set $ \beta = -6 $, similar to \cite{Popchev2015,Ramazanoglu2016,Yazadjiev2016, StaykovPopchev2018}, as it is a moderate value for which big enough deviation from GR is observed. We will use massless STT, $ (\beta = -6, m_\varphi = 0, \lambda = 0 ) $ and STT with massive scalar field with dimensionless masses $ (\beta = -6, m_{\varphi} = \{5\times 10^{-3}, 5 \times 10^{-2}\}, \lambda = 0 ) $, which fall well within the range given by eq. (\ref{STT_m_interval}) and for the chosen $ \beta $ present very well the influence of the mass. We will use massless STT with self-interaction with dimensionless set of values $ (\beta = -6, m_\varphi = 0, \lambda = \{ 0.1, 1, 10 \} ) $ and STT with massive self-interacting scalar field with set of values $(\beta = -6, m_{\varphi} = \{5\times 10^{-3}, 5 \times 10^{-2}\}, \lambda = \{ 0.1, 1, 10 \} ) $. 
    
    \paragraph*{} In this study we are using the piecewise polytropic approximation of several EOS \cite{Read2009}. They are chosen in such a way in order to cover a wide variety of theoretical approaches such as nuclear many body approach, e.g. APR, relativistic mean field theory approach, e.g. SLy, and others. Similar to \cite{Staykov2016}, we also include softer EOS already excluded by the observations of two solar mass neutron stars because in STT under certain initial conditions the maximum mass can reach, even exceed, this observational limit due to scalarization effect. In addition, considering a very wide set of EOS, even if they are excluded from the observations, can give us a more profound insight whether the observed EOS universality is due to the limited set of chosen EOS or is indeed an intrinsic feature of the theory.
    
    \begin{figure}[t]
        \includegraphics[width=0.45\textwidth]{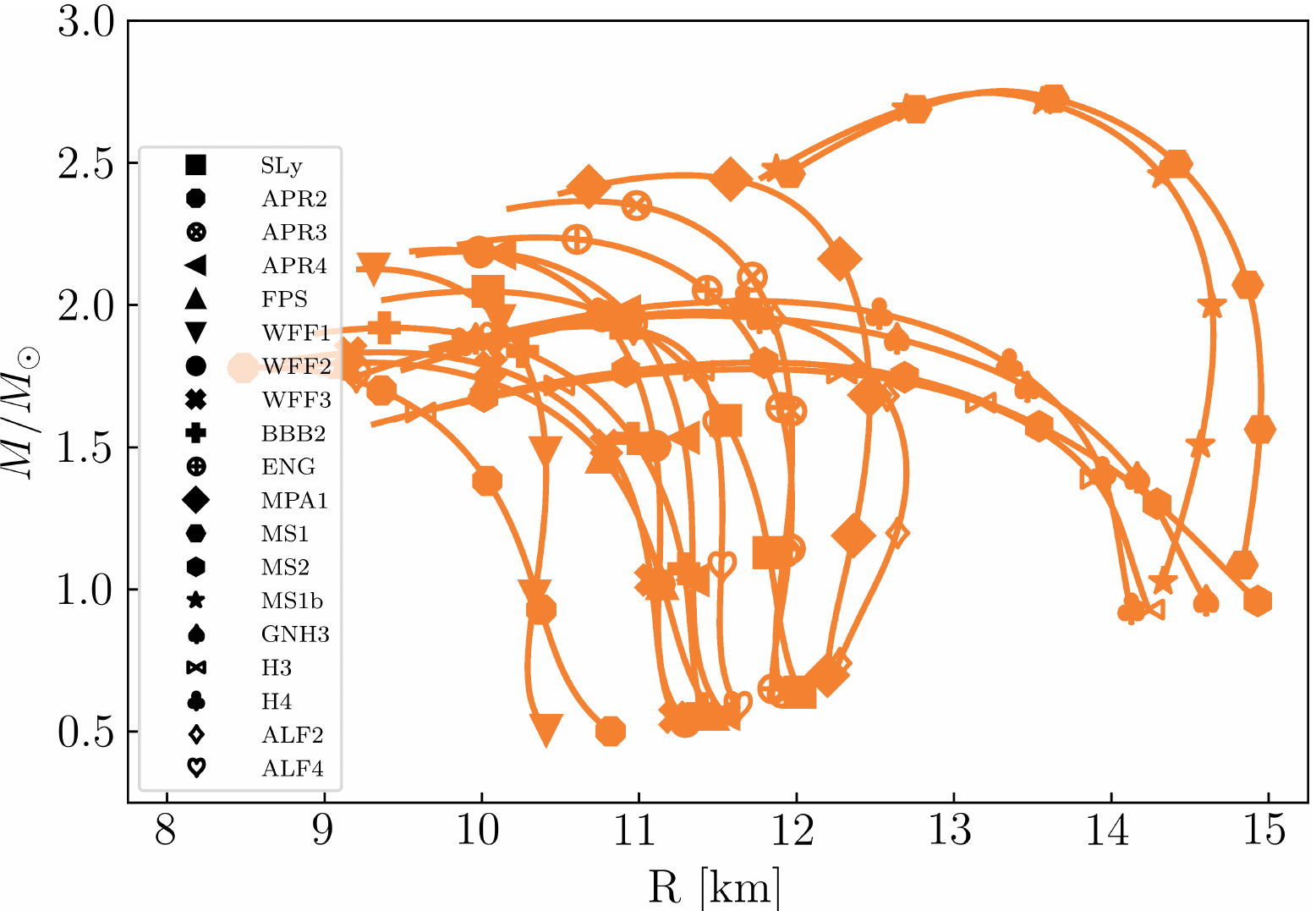}
        \includegraphics[width=0.45\textwidth]{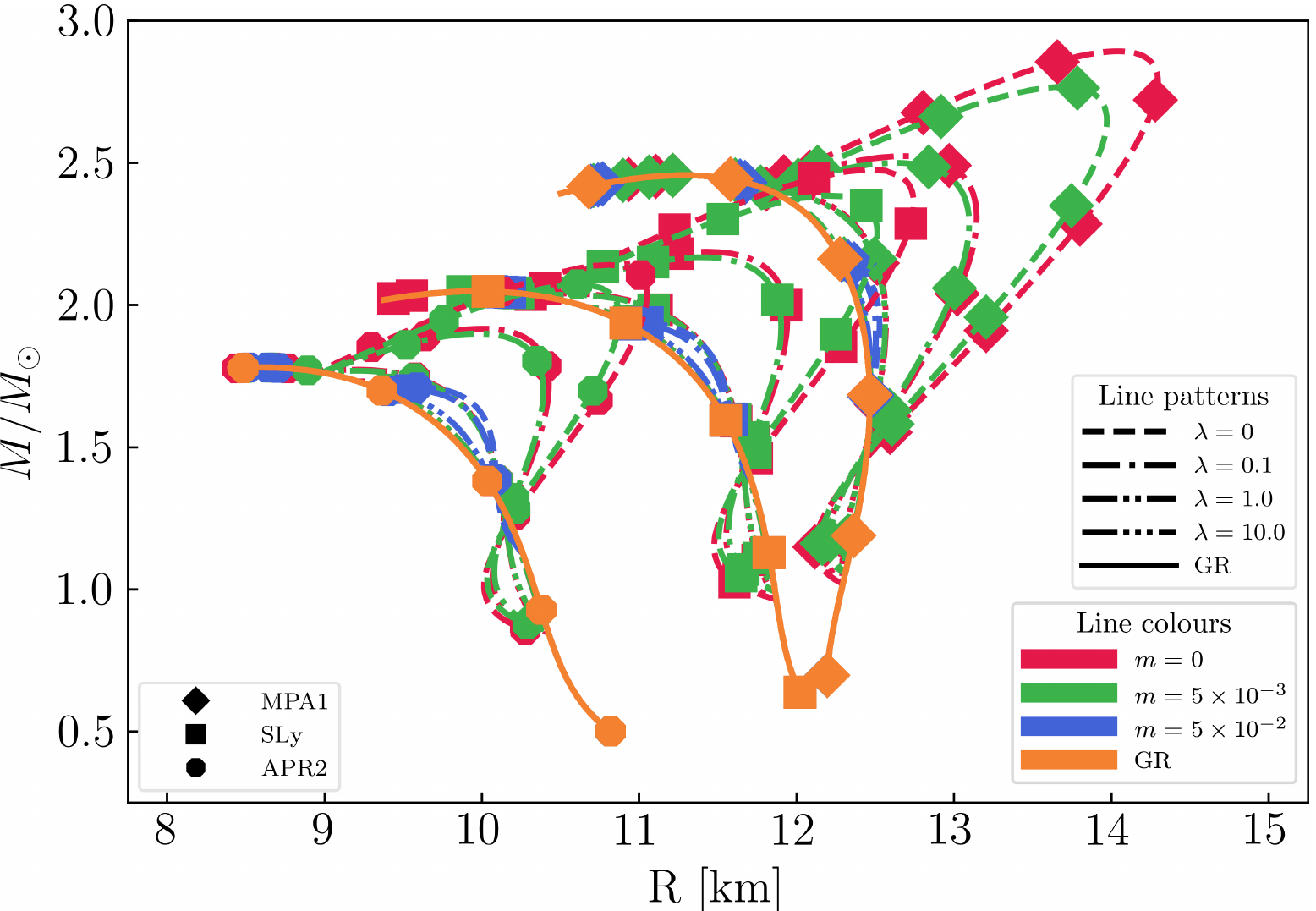}
        \caption{
            Mass of radius relation for all of the employed EOS in GR \textit{(left panel)}, and representative EOS -- APR2, MPA1, SLy, in GR and in STT with fixed $ \beta = -6 $ and various values of $ m_{\varphi} $ and $ \lambda $ \textit{(right panel)}. On both figures unique marker is assigned to each EOS and GR results are presented with \GetGRc{} \GetGRls{}.
            \textit{Right}: STT solutions with scalar field mass are presented with different colour: $ m_{\varphi} = 0 $ -- \GetSTTmZero{}, $ m_{\varphi} = 5\times10^{-3} $ -- \GetSTTmMthree{}, $ m_{\varphi} = 5\times 10^{-2} $ -- \GetSTTmMtwo{}. For each mass, solutions for various self-interactions are presented with different line patterns: $ \lambda = 0 $ -- \GetSTTlambdaZero{}, $ \lambda = 0.1 $ -- \GetSTTlambdaMMone{}, $ \lambda = 1 $ -- \GetSTTlambdaMzero, $ \lambda = 10 $ -- \GetSTTlambdaMPone.
        }
        \label{MvsR_all} 
    \end{figure}
    
    \paragraph*{} In Fig. \ref{MvsR_all} we are plotting two mass-radius-relations: in the left pane all the EOS employed in the present study are shown  for pure GR only; in the right panel only representative EOS are used -- stiff (MPA1), moderate (SLy) and soft(APR2), and the results are both for GR and STT. As one can see, indeed there is a degeneracy between effects coming from the presence of scalar field or from varying the EOS. In the right panel, the effects of the mass and self-interaction are also clearly visible. Although both of them independently suppress scalarization, the $ \sim \varphi^{4} $ term retains the position of the bifurcation points of the massless STT without self-interaction (\GetSTTmZero{} \GetSTTlambdaZero{}), while the $ \sim \varphi^{2} $ decreases the distance between them. The latter means that even for big values for $ \lambda $, i.e. highly suppressed scalarization, we will have a wider range of central densities for which scalarization can occur contrary to massive case. Further discussion can be found in \cite{StaykovPopchev2018}.
    
    \subsection{Normalized moment of inertia -- compactness relations}
    
    \paragraph*{} Here we will investigative the universality in normalized moment of inertia -- compactness relations, suggested for the first time in \cite{Ravenhall1994}, and extensively studied by Breu and Rezzolla \cite{Breu2016} in GR and Staykov et al. \cite{Staykov2016} in $ f(R) $ gravity and massless STT.
    
\begin{figure}[t]
        \includegraphics[width=0.8\textwidth]{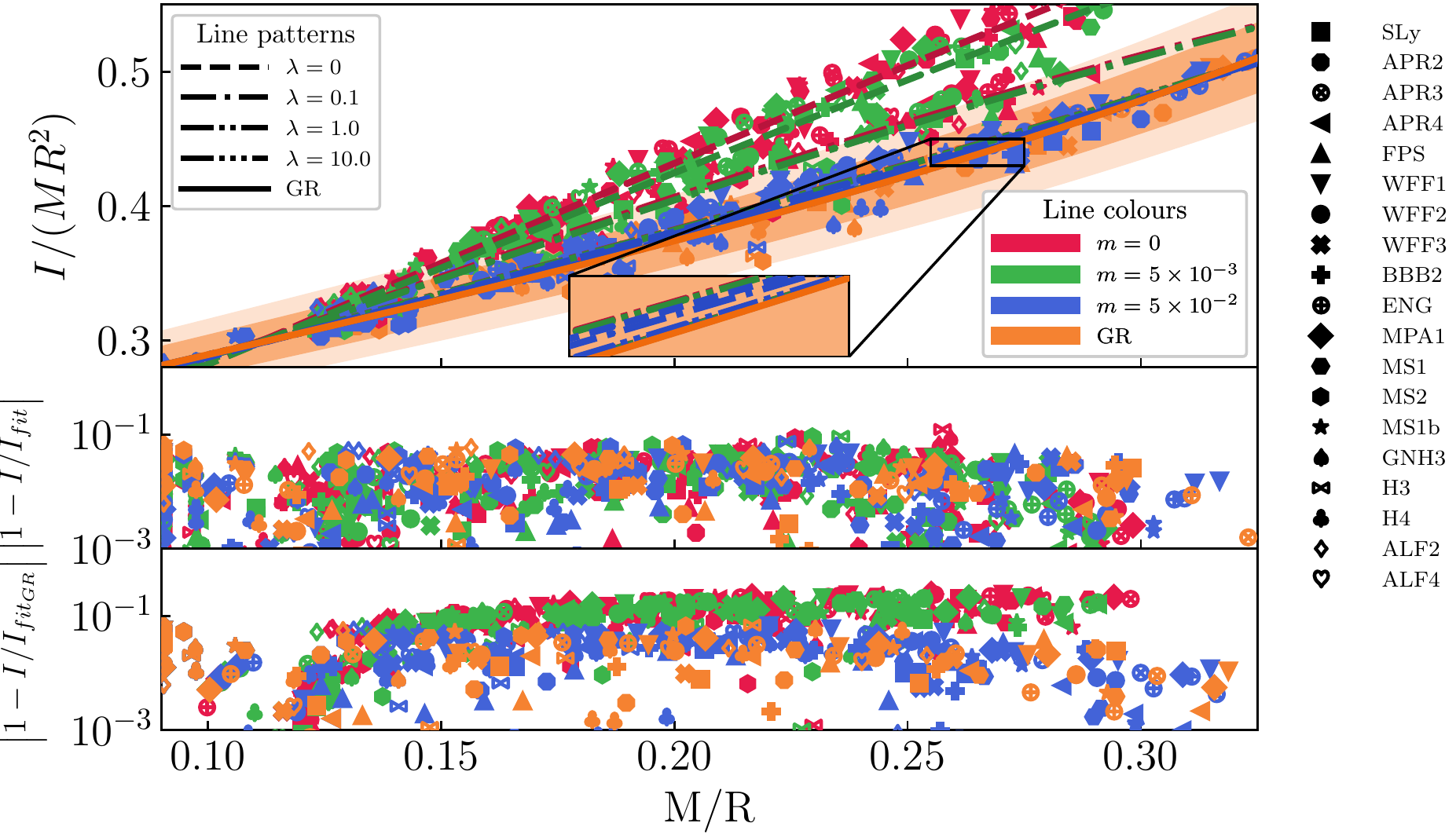}
        \caption{ 
            The normalized moment of inertia, $ I/MR^2 $, as a function of the stellar compactness $ M/R $ in GR and STT with $ \beta = -6 $ and various values of $ m_\varphi $ and $ \lambda $, with corresponding polynomial fits, are shown. EOS are indicated with individual symbol, GR results are represented with \GetGRc{} and models in STT with different masses of the scalar field are presented with different colour: $ m_{\varphi} = 0 $ -- \GetSTTmZero{}, $ m_{\varphi} = 5\times10^{-3} $ -- \GetSTTmMthree{}, $ m_{\varphi} = 5\times 10^{-2} $ -- \GetSTTmMtwo{}. For all masses, results for various self-interaction terms are calculated and the corresponding polynomial fits are presented with different line pattern: $ \lambda = 0 $ -- \GetSTTlambdaZero{}, $ \lambda = 0.1 $ -- \GetSTTlambdaMMone{}, $ \lambda = 1 $ -- \GetSTTlambdaMzero, $ \lambda = 10 $ -- \GetSTTlambdaMPone. In the middle panel the corresponding deviations of the polynomial fits and the data $ \left| 1 - \tilde I/\tilde I_{\mbox{fit}}\right| $ are presented while in the bottom panel -- the deviations from the GR polynomial fit $ \left| 1 - \tilde I/\tilde I_{\mbox{fit GR}}\right| $ for all data, including the STT models.}
        \label{Fig_Uni_tilde}
    \end{figure}

    \paragraph*{} In Fig. \ref{Fig_Uni_tilde} we are plotting the  normalized moment of inertia $ \widetilde I \equiv I/(MR^{3}) $ as a function of compactness $ M/R $ in GR and in STT with fixed $ \beta = -6 $ and various sets of values $ (m_{\varphi}, \lambda) $.  The results show quite good EOS universality for fixed $ (m_{\varphi}, \lambda) $. For large compactnesses $ M/R $ the deviations from GR, due to scalarization effect, are larger than the EOS uncertainty, e.g. NS for massive STT with compactness $ M/R \gtrsim 0.25 $ separate from massive STT with self-interaction $ \lambda = 0.1 $ in the same interval. 
    
    \paragraph*{} We are fitting the different theories separately with polynomial fit of forth order, with excluded second and third order terms having the following form
    \begin{eqnarray}
    \label{tilde_I} && \widetilde I = \widetilde a_{0} + \widetilde a_{1} \dfrac{M}{R} + \widetilde a_{4} \left(\dfrac{M}{R}\right)^{4},
    \end{eqnarray}
    which gives small correction to the natural choice -- the linear fit. This form of the fit is suggested for the first time by Lattimer and Schutz \cite{Lattimer2005},  studied further by Breu and Rezzolla \cite{Breu2016} and later used by Staykov et al. \cite{Staykov2016} in alternative theories of gravity.
    
    \paragraph*{} The numerical values of the fitting coefficients and the corresponding $ \chi^{2} $ estimations of each fit in Fig. \ref{Fig_Uni_tilde}  can be found in Table \ref{Table_uni_coeff_tildeI} in the Appendix. In the middle panel of the graph we are plotting the relative deviation of the data points from the fitting curve for each theory. We define the deviations as $ \left| 1 - \widetilde I/\widetilde I_{\mbox{fit}}\right| $ and for the set of EOS we are using it is below $ 10\% $ for all theories studied here.
    
    \begin{figure}[h]
        \includegraphics[width=0.45\textwidth]{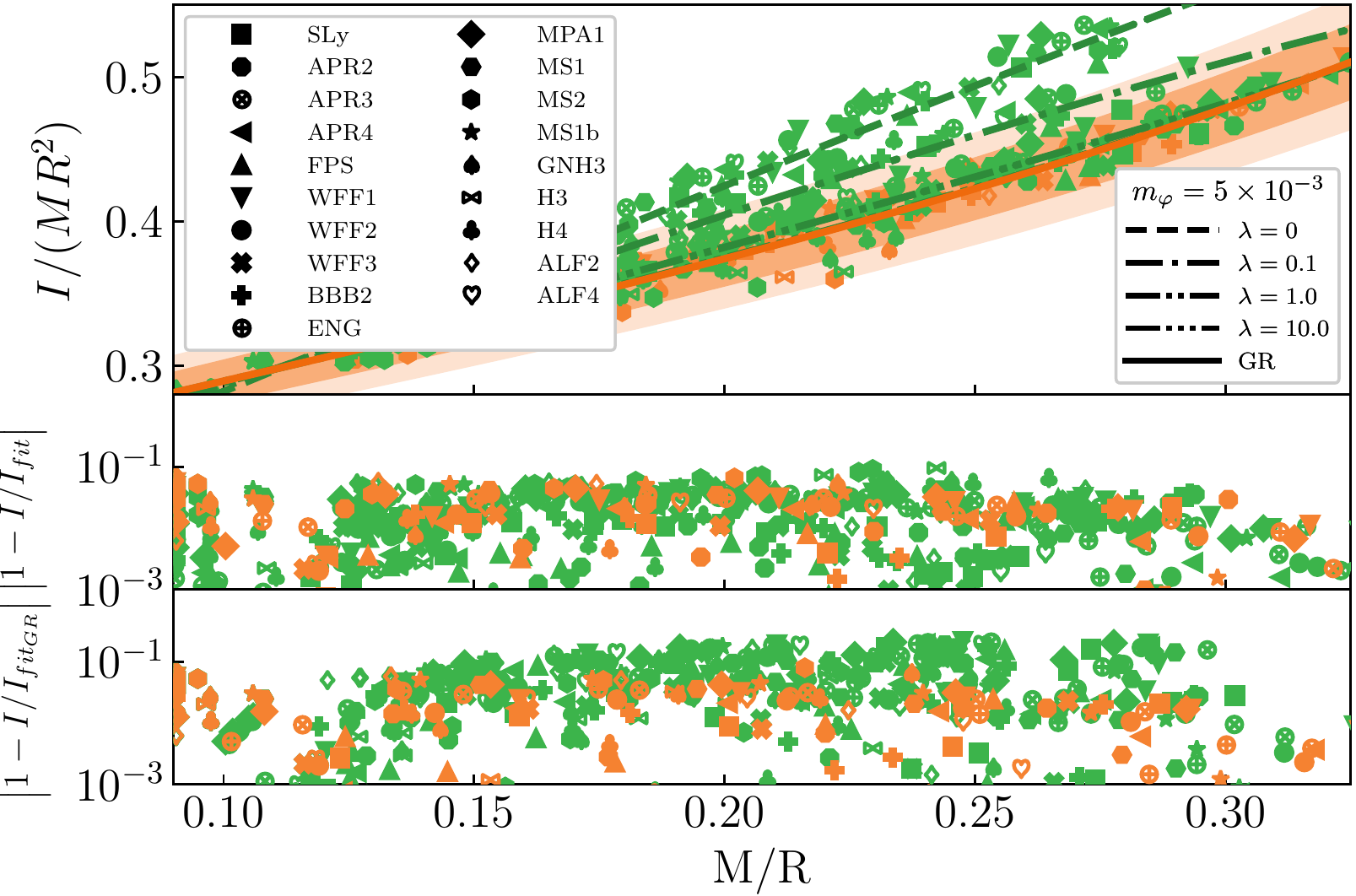}
        \includegraphics[width=0.45\textwidth]{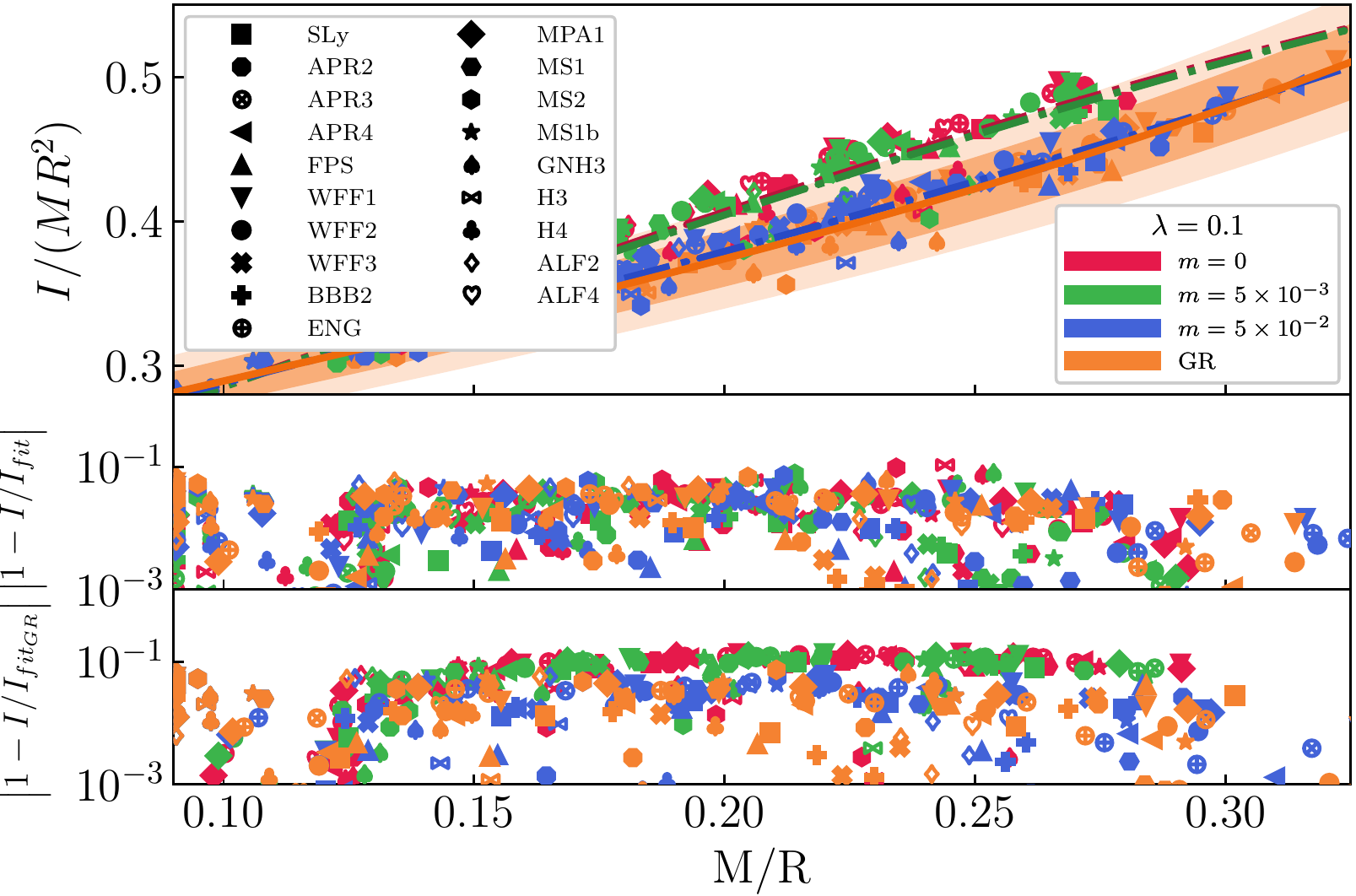}
        \caption{  
            The normalized moment of inertia $ I/MR^2 $ as a function of the stellar compactness $ M/R $. (\textit{Left panel}) Neutron star models in massive STT with self-interaction with fixed $ m = 5\times 10^{-3} $ and several different values for $ \lambda $ are plotted using \GetSTTmMthree{} line in different patterns. (\textit{right panel}) Neutron star models in massive STT with self-interaction with fixed $ \lambda = 0.1 $  and several different values of $ m_{\varphi} $ are plotted using \GetSTTlambdaMMone{} in different colours. In both figures the corresponding polynomial fits to the data of the form given by eq. (\ref{tilde_I}) are also presented. In the middle of each graph the corresponding deviations of the polynomial fits and the data $ \left| 1 - \tilde I/\tilde I_{\mbox{fit}}\right| $ are presented while in the bottom -- the deviations from the GR polynomial fit $ \left| 1 - \tilde I/\tilde I_{\mbox{fit GR}}\right| $ for all data, including the STT models.
        } 
        \label{Fig_UniTilde_m5e-3_lambda1e-1}
    \end{figure}
    
    \paragraph*{} Additionally we present several residual norms for each fit -- $ <L> $, the average over all EOSs of all the residuals $ \left| 1 - \widetilde I/\widetilde I_{\mbox{fit}}\right| $; $ <L_{\infty}> $, the average over all EOSs of the largest relative deviation between the data and fit; $ L_{\infty} $, the largest residual across all EOSs. A summary of the values of the various norms for both STT and GR is shown in Table \ref{Table_uni_resd_tildeI} in the Appendix. The $ <L_{\infty}> $ and $ L_{\infty} $ residual norms of the GR fit are presented in the plot as shaded areas, with the latter one being with lighter colour than the former. 
    
    \begin{figure}[t]
        \includegraphics[width=0.8\textwidth]{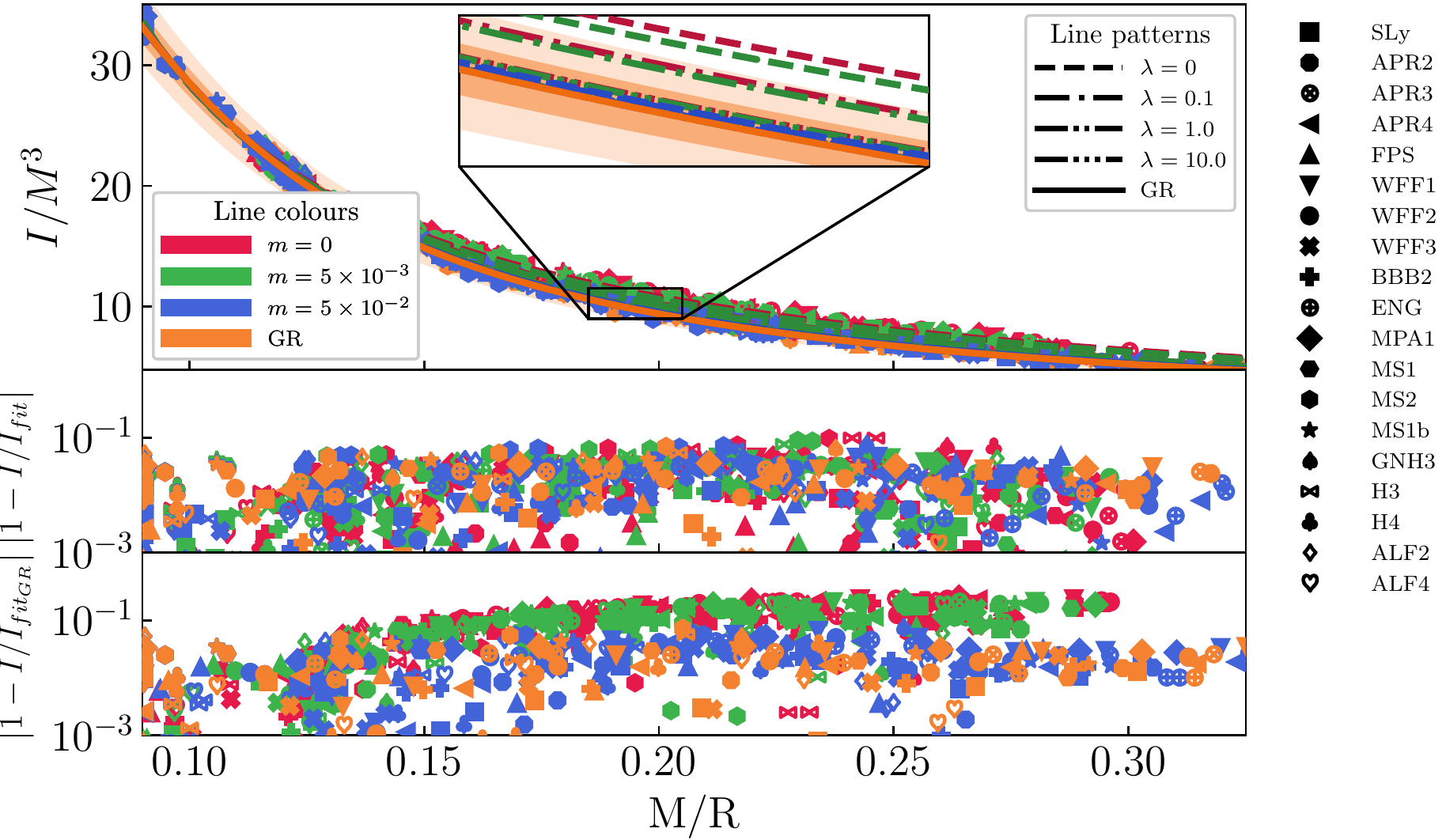}
        \caption{   
            The normalized moment of inertia $ I/M^3 $  as a function of the stellar compactness $ M/R $ in GR and STT with $ \beta = -6 $ and various values of $ m_\varphi $ and $ \lambda $, with corresponding polynomial fits are shown. The notations are same as in Fig. \ref{Fig_Uni_tilde}. }
        \label{Fig_Uni_bar}
    \end{figure}
    
    \begin{figure}[h]
        \includegraphics[width=0.45\textwidth]{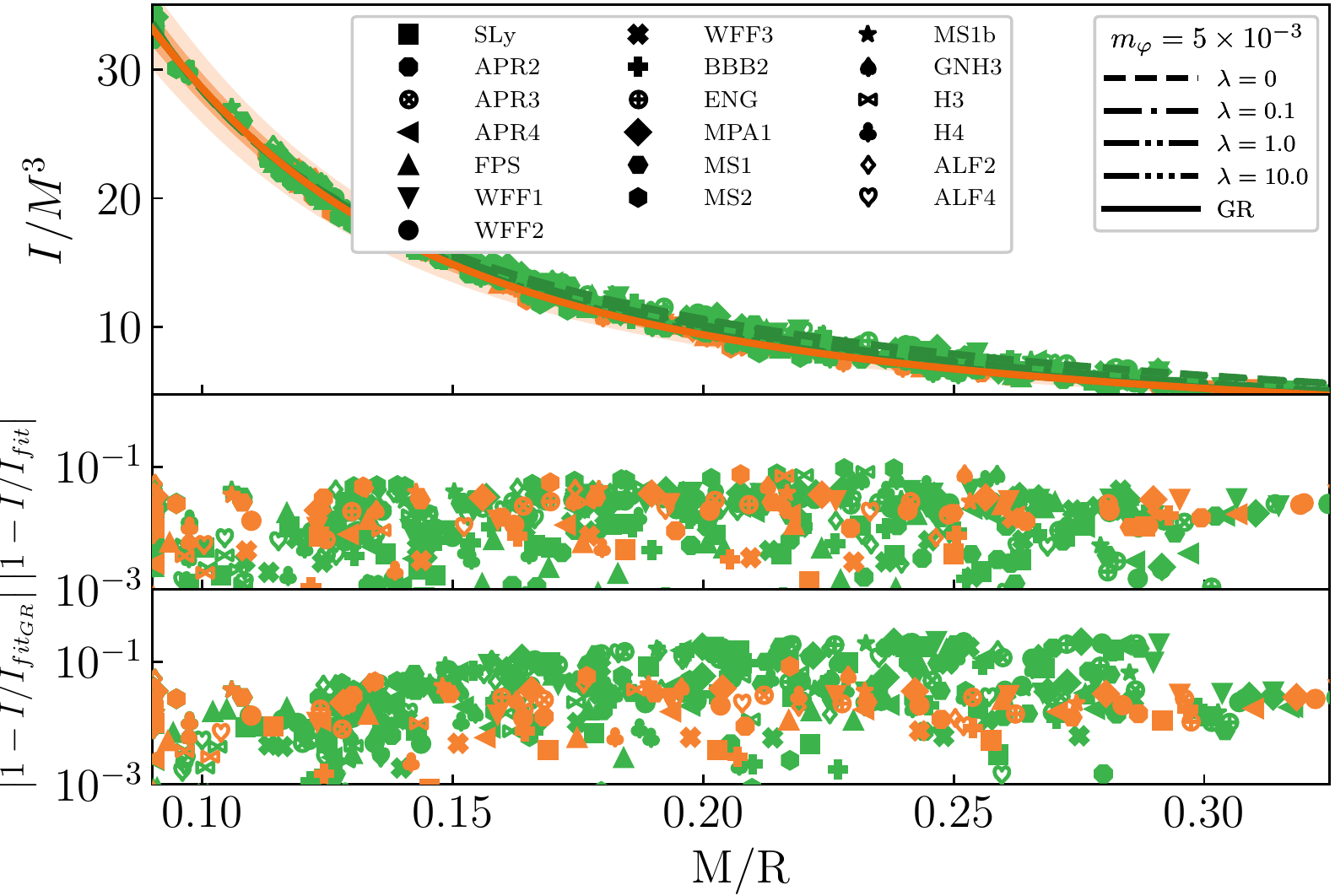}
        \includegraphics[width=0.45\textwidth]{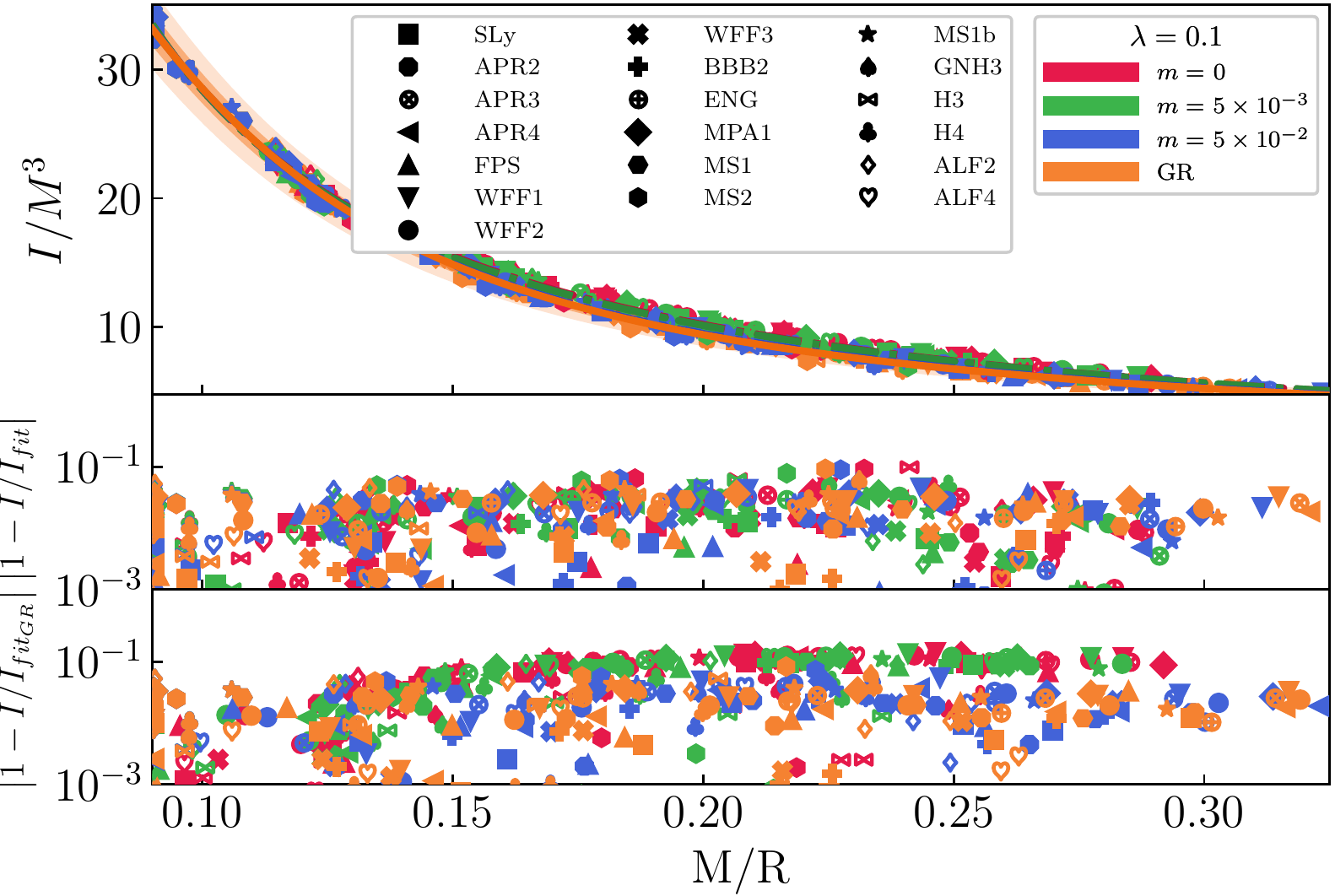}
        \caption{  
            The normalized moment of inertia $ I/M^3 $ as a function of the stellar compactness $ M/R $ are plotted. The notations are the same as in Fig. \ref{Fig_UniTilde_m5e-3_lambda1e-1}.
        } 
        \label{Fig_UniBar_m5e-3_lambda1e-1}
    \end{figure}

    \paragraph*{} In Fig \ref{Fig_UniTilde_m5e-3_lambda1e-1} we present a simplified version of Fig. \ref{Fig_Uni_tilde} by fixing one additional free parameter -- either $ m_{\varphi} = 5\times10^{-3} $ in the left panel or $ \lambda = 0.1 $ in the right panel, and plotting the results when varying the corresponding remaining one. On both figures we observe that when either $ m_{\varphi} $ or $ \lambda $ is increased the scalarization effect is suppressed and the results for massive STT with self-interaction fall well within the fit uncertainty of GR and even converge to it. 
    
    \paragraph*{} In Fig. \ref{Fig_Uni_bar} we are using different normalization for the moment of inertia, namely $ \overline I \equiv I/M^{3} $. Qualitative this choice for normalization is different from $ \widetilde I $, as with the increase of compactness $ \overline I $ decreases, but otherwise we observe similar behaviour -- all the results show quite good universality. For small compactness both STT and GR cluster together and for  large compactness (roughly $ M/R \gtrsim 0.25 $) the scalarized results deviate non-negligibly from the GR ones. We are using polynomial of the form:
    \begin{eqnarray}
    \label{bar_I} && \overline I = \overline a_{1} \left(\dfrac{M}{R}\right)^{-1} + \overline a_{2} \left(\dfrac{M}{R}\right)^{-2} + \overline a_{3} \left(\dfrac{M}{R}\right)^{-3} + \overline a_{4} \left(\dfrac{M}{R}\right)^{-4}
    \end{eqnarray}
    to fit each of the presented cases separately. Visually $ \overline I $ looks like the better normalization, compared to $ \widetilde I $, but in fact this is a misleading artifact of the plot due to the different scales used on the $y$-axis. Indeed, if one compares the data in the lower plots, which again is the deviation between fit and data, he will find out that for the same set of EOSs it does not exceed $ 10\% $. Even more, the residual norms for each normalization (for $ \widetilde I $ are in Table \ref{Table_uni_resd_tildeI} and for $ \overline I $ are in Table \ref{Table_uni_resd_barI}) are comparable: the typical average over all EOSs of all residuals per theory is $ <L> \lesssim 2\% $; the average over all EOSs of maximum residual per theory is $ <L_{\infty}> \lesssim 5\% $; largest residual across all EOS per theory is $ L_{\infty} \lesssim 10\% $, with the softest (larger compactness $ M/R $) EOS being the major contributors to the largest relative deviations between data and fit. 
    
    \paragraph*{} We also include simplified version of Fig. \ref{Fig_Uni_bar} that is Fig. \ref{Fig_UniBar_m5e-3_lambda1e-1}, where the free parameters -- $ m_{\varphi} = 5\times10^{-3} $\textit{(left)} and $ \lambda = 0.1 $\textit{(right)}, are fixed. Qualitative, as in Fig.  \ref{Fig_UniTilde_m5e-3_lambda1e-1}, the massive STT theories with self-interaction is well enough separated from GR for large compactness and small ($ m_{\varphi}, \lambda $), but with the increase $ m_{\varphi} $ or $ \lambda $ they converge to GR.
    
    \paragraph*{} Let us now comment on the differences with the GR case. In each of the Figs. \ref{Fig_Uni_tilde}--\ref{Fig_UniBar_m5e-3_lambda1e-1} in the bottom panel the relative deviation from the GR fit is shown. As one can see, the deviations of the scalarized models are clearly larger than the equation of state uncertainty reaching up to 20\%. We should note, that this is only for the moderate chosen values of the parameters. For even smaller $\beta$ these deviations can increase considerably. That is why the universal relations can be potentially used to constrain observationally the massive STT with self-interaction. On the other hand, as the intuition from building equilibrium models also shows \cite{StaykovPopchev2018}, even for very small $\beta$ there exist a range of parameters, typically for higher $\lambda$, where the models are still scalarized for a large range of central energy densities, but the properties of the neutron stars and the corresponding universal relations are almost indistinguishable for GR.
        
    \section{Conclusion}
    
    \paragraph*{} The study of universal relations between NS parameters is appealing for the prospects to provide us with powerful tool to determine difficult to obtain parameters of the stars. For example, while the mass and the moment of inertia can be measured with good accuracy, the optical measurement of the radius depend on different factors, like the redshift, the distance to the star, the effect of the atmosphere, the absorption in the interstellar space, etc. \cite{Lattimer2005}. Such relations can helps us overcome the large uncertainty in the EOS and allows us to establish uniquely the star parameters, and as a result help us to accurately test alternative theories of gravity against GR.
    
    \paragraph*{} The current state-of-the-art observations of $ M, I \mbox{ and } R $ make them the natural choice for forming universal relations, as they can be measured with good accuracy for binary systems. Our study shows that using $ I/(MR^{2}) $ or $ I/M^{3} $ as functions of compactness leads to quite good universality in agreement with \cite{Staykov2016, Breu2016} not only for GR, but for all examined classes of STT, and for our set of EOS. The deviations, per theory, is not higher than $ 10 \% $ for both studied normalisations, with the largest deviations being for the softer EOS. The differences between the universal relations for scalarized models and the GR ones is clearly larger than the spread of the data due to the EOS uncertainty and it reaches up to roughly 20\% for the considered values of $\beta=-6$. The difference will naturally increase with the decrease of $\beta$. This shows that the considered relations can serve as way to constraint the massive STT with self interaction independently of the EOS uncertainty. 
    
    \paragraph*{} Among the studied subclasses of STT, particularly interesting is the massive case, $ (\beta = -6, m_{\varphi} = 5\times10^{-3}, \lambda = 0) $, for which we know that the maximum deviation for the moment of inertia can be of order $ 40\% $, for the EOS studied in \cite{Yazadjiev2016}, but for each of the examined normalized relations the deviation from GR is about $ 10\% $ for the whole interval of compactnesses. We should stress out that although values $ \beta \lesssim -4.5 $ for massless STT are restricted by the astronomical observations, this is not the case when investigating massive STT for which $ (\beta = -6, m_{\varphi} = 5\times10^{-3}, \lambda = 0) $ is well within the allowed interval. Moreover, massive STT with self-interaction with large $ \lambda $ admit greater $ \beta < 0 $ with simultaneously exhibiting smaller deviation from GR while scalarization occurs for a large range of central energy densities. Hence one can say that these relations are not only EOS independent, but for a large part of the domain of the chosen STT, theory independent too.
    
    \section{Acknowledgements}
    
    \paragraph*{} KS, SY, and DD would like to thank for support by the COST Actions CA15117, CA16104 and CA16214. SY, KS and DP would like to thank for the support by the Bulgarian NSF Grant DCOST 01/6. SY is supported partially by the Sofia University Grant No 3258/2017. DD would like to thank the European Social Fund, the Ministry of Science, Research and the Arts Baden-W\"urttemberg for the support. DD is indebted to the Baden-W\"urttemberg Stiftung for the financial support of this research project by the Eliteprogramme for Postdocs.
    
    \bibliography{references}
    
    \clearpage
    \section{Appendix}
    In the appendix we will give tables with detailed information about the fits used in the figures.
    \begin{table}[h]
        \makebox[\textwidth][c]{ \begin{tabular}{ c | c | c | c | c }
                \backslashbox{m}{$ \lambda $} & $ 0 $ & $ 0.1 $ & $ 1 $ & $ 10 $\\
                \hline
                $ 0 $ & 
                $ \begin{aligned}[c]
                a_0 &= 0.131 \\
                a_1 &= 1.514 \\
                a_4 &= -1.263 \\
                \chi_r^2 &= 6.044\times10^{-5} \\
                \end{aligned} $ & 
                $ \begin{aligned}[c]
                a_0 &= 0.160 \\
                a_1 &= 1.255 \\
                a_4 &= -3.061 \\
                \chi_r^2 &= 5.461\times10^{-5} \\
                \end{aligned} $ & 
                $ \begin{aligned}[c]
                a_0 &= 0.196 \\
                a_1 &= 0.921 \\
                a_4 &= 1.136 \\
                \chi_r^2 &= 4.764\times10^{-5} \\
                \end{aligned} $ &
                $ \begin{aligned}[c]
                a_0 &= 0.209 \\
                a_1 &= 0.800 \\
                a_4 &= 3.662 \\
                \chi_r^2 &= 4.189\times10^{-5} \\
                \end{aligned} $ \\
                \hline
                $ 5 \times 10^{-3} $ & 
                $ \begin{aligned}[c]
                a_0 &= 0.137 \\
                a_1 &= 1.445 \\
                a_4 &= -1.331 \\
                \chi_r^2 &= 5.703\times10^{-5} \\
                \end{aligned} $ & 
                $ \begin{aligned}[c]
                a_0 &= 0.162 \\
                a_1 &= 1.230 \\
                a_4 &= -2.621 \\
                \chi_r^2 &= 5.329\times10^{-5} \\
                \end{aligned} $ & 
                $ \begin{aligned}[c]
                a_0 &= 0.196 \\
                a_1 &= 0.918 \\
                a_4 &= 1.215 \\
                \chi_r^2 &= 4.724\times10^{-5} \\
                \end{aligned} $ & 
                $ \begin{aligned}[c]
                a_0 &= 0.209 \\
                a_1 &= 0.798 \\
                a_4 &= 3.720 \\
                \chi_r^2 &= 4.270\times10^{-5} \\ 
                \end{aligned} $ \\
                \hline
                $ 5 \times 10^{-2} $ & 
                $ \begin{aligned}[c]
                a_0 &= 0.195 \\
                a_1 &= 0.917 \\
                a_4 &= 1.122 \\
                \chi_r^2 &= 4.211\times10^{-5}\\ 
                \end{aligned} $ & 
                $ \begin{aligned}[c]
                a_0 &= 0.199 \\
                a_1 &= 0.882 \\
                a_4 &= 1.862 \\
                \chi_r^2 &= 4.402\times10^{-5} \\ 
                \end{aligned} $ & 
                $ \begin{aligned}[c]
                a_0 &= 0.207 \\
                a_1 &= 0.817 \\
                a_4 &= 3.280 \\
                \chi_r^2 &= 4.321\times10^{-5} \\
                \end{aligned} $ &
                $ \begin{aligned}[c]
                a_0 &= 0.210 \\
                a_1 &= 0.789 \\
                a_4 &= 3.905 \\
                \chi_r^2 &= 4.193\times10^{-5} \\
                \end{aligned} $ \\
                \hline
                \multicolumn{5}{c}{$ \mbox{GR:}
                    a_0 = 2.103e-01;
                    a_1 = 7.877e-01;
                    a_4 = 3.950e+00;
                    \chi_r^2 = 4.186e-05
                    $}\\
                \hline
        \end{tabular} }
        \caption{ { \footnotesize $ I/MR^2 $ fit coefficients and corresponding $ \chi^{2} $ values. The first row and column hold all the presented in the paper values for $ m_{\varphi} $ and $ \lambda $. The entries in each cell present the numerical value of coefficients in \eqref{tilde_I} and the corresponding $ \chi^{2} $ } }
        \label{Table_uni_coeff_tildeI}
    \end{table}
    
    \begin{table}
        \makebox[\textwidth][c]{ \begin{tabular}{ c | c | c | c | c }
                \backslashbox{m}{$ \lambda $} & $ 0 $ & $ 0.1 $ & $ 1 $ & $ 10 $\\
                \hline
                $ 0 $ & 
                $ \begin{aligned}[c]
                <&L> = 2.131\times10^{-2} \\
                <&L_\infty> = 5.249\times10^{-2} \\
                &L_\infty = 1.228\times10^{-1} \\
                \end{aligned} $ & 
                $ \begin{aligned}[c]
                <&L> = 2.131\times10^{-2} \\
                <&L_\infty> = 5.249\times10^{-2} \\
                &L_\infty = 1.228\times10^{-1} \\
                \end{aligned} $ & 
                $ \begin{aligned}[c]
                <&L> = 2.268 \times 10^{-2} \\
                <&L_{\infty}> = 5.020 \times 10^{-2} \\ 
                &L_{\infty} = 1.018 \times 10^{-1}\\
                \end{aligned} $ &
                $ \begin{aligned}[c]
                <&L> = 2.216 \times 10^{-2} \\
                <&L_{\infty}> = 5.198 \times 10^{-2} \\ 
                &L_{\infty} = 9.286 \times 10^{-2}\\
                \end{aligned} $\\ 
                \hline
                $ 5 \times 10^{-3} $ & 
                $ \begin{aligned}[c]
                <&L> = 2.171 \times 10^{-2} \\
                <&L_{\infty}> = 5.814 \times 10^{-2} \\ 
                &L_{\infty} = 1.257 \times 10^{-1}\\
                \end{aligned} $ & 
                $ \begin{aligned}[c]
                <&L> = 2.238 \times 10^{-2} \\
                <&L_{\infty}> = 4.874 \times 10^{-2} \\ 
                &L_{\infty} = 1.149 \times 10^{-1}\\
                \end{aligned} $ & 
                $ \begin{aligned}[c]
                <&L> = 2.259 \times 10^{-2} \\
                <&L_{\infty}> = 4.982 \times 10^{-2} \\ 
                &L_{\infty} = 9.963 \times 10^{-2}\\
                \end{aligned} $ & 
                $ \begin{aligned}[c]
                <&L> = 2.241 \times 10^{-2} \\
                <&L_{\infty}> = 5.213 \times 10^{-2} \\ 
                &L_{\infty} = 9.271 \times 10^{-2}\\ 
                \end{aligned} $ \\
                \hline
                $ 5 \times 10^{-2} $ & 
                $ \begin{aligned}[c]
                <&L> = 2.143 \times 10^{-2} \\
                <&L_{\infty}> = 4.796 \times 10^{-2} \\ 
                &L_{\infty} = 9.661 \times 10^{-2}\\
                \end{aligned} $ & 
                $ \begin{aligned}[c]
                <&L> = 2.198 \times 10^{-2} \\
                <&L_{\infty}> = 4.920 \times 10^{-2} \\ 
                &L_{\infty} = 9.481 \times 10^{-2}\\
                \end{aligned} $ & 
                $ \begin{aligned}[c]
                <&L> = 2.222 \times 10^{-2} \\
                <&L_{\infty}> = 5.119 \times 10^{-2} \\ 
                &L_{\infty} = 9.295 \times 10^{-2}\\
                \end{aligned} $ &
                $ \begin{aligned}[c]
                <&L> = 2.235 \times 10^{-2} \\
                <&L_{\infty}> = 5.205 \times 10^{-2} \\ 
                &L_{\infty} = 9.246 \times 10^{-2}\\
                \end{aligned} $ \\
                \hline
                \multicolumn{5}{c}{$ \mbox{GR:}
                    <L> = 2.236 \times 10^{-2},
                    <L_{\infty}> = 5.209 \times 10^{-2}, 
                    L_{\infty} = 9.262 \times 10^{-2},
                    $}\\
                \hline
        \end{tabular} }
        \caption{ { \footnotesize Summary of various averaged norms of $ I/MR^2 $ fit residuals $ \left| 1 - \tilde I/\tilde I_{\mbox{fit}}\right| $. The first row and column hold all the presented in the paper values for $ m_{\varphi} $ and $ \lambda $. The entries in each cell present each fit as follows - $ <L_{1}> $, the average over all EOSs of all the residuals; $ <L_{\infty}> $, the average over all EOSs of the largest relative deviation between the data and fit; $ L_{\infty} $, the largest residual across all EOSs. Only the GR $ <L_{\infty}> $ and $ L_{\infty} $ are included in Fig.  \ref{Fig_Uni_tilde} as shaded areas.} }
        \label{Table_uni_resd_tildeI}
    \end{table}
    
    \begin{table}
        \makebox[\textwidth][c]{ \begin{tabular}{ c | c | c | c | c }
                \backslashbox{m}{$ \lambda $} & $ 0 $ & $ 1 \times 10^{-1} $ & $ 1 \times 10^{0} $ & $ 1 \times 10^{1} $\\
                \hline
                $ 0 $ & 
                $ \begin{aligned}[c]
                a_1 &= 1.175 \\
                a_2 &= 0.282 \\ 
                a_3 &= -2.267 \times 10^{-2} \\
                a_4 &= 1.099 \times 10^{-3} \\
                \chi_{r}^2 &= 0.154 \\
                \end{aligned} $ & 
                $ \begin{aligned}[c]
                a_1 &= 0.763 \\
                a_2 &= 0.343 \\ 
                a_3 &= -2.182 \times 10^{-2} \\
                a_4 &= 8.227 \times 10^{-4} \\
                \chi_{r}^2 &= 0.148 \\
                \end{aligned} $ & 
                $ \begin{aligned}[c]
                a_1 &= 1.052 \\
                a_2 &= 0.121 \\ 
                a_3 &= 1.500 \times 10^{-2} \\
                a_4 &= -9.189 \times 10^{-4} \\
                \chi_{r}^2 &= 0.127 \\
                \end{aligned} $ &
                $ \begin{aligned}[c]
                a_1 &= 1.253 \\
                a_2 &= 8.055 \times 10^{-3} \\ 
                a_3 &= 3.207 \times 10^{-2} \\
                a_4 &= -1.689 \times 10^{-3} \\
                \chi_{r}^2 &= 0.121 \\
                \end{aligned} $ \\
                \hline
                $ 5 \times 10^{-3} $ & 
                $ \begin{aligned}[c]
                a_1 &= 1.206 \\
                a_2 &= 0.242 \\ 
                a_3 &= -1.582 \times 10^{-2} \\ 
                a_4 &= 7.863 \times 10^{-4} \\ 
                \chi_{r}^2 &= 0.152 \\
                \end{aligned} $ & 
                $ \begin{aligned}[c]
                a_1 &= 0.853 \\
                a_2 &= 0.298 \\ 
                a_3 &= -1.584 \times 10^{-2} \\ 
                a_4 &= 5.805 \times 10^{-4} \\ 
                \chi_{r}^2 &= 0.147 \\
                \end{aligned} $ & 
                $ \begin{aligned}[c]
                a_1 &= 1.069 \\
                a_2 &= 0.113 \\ 
                a_3 &= 1.603 \times 10^{-2} \\
                a_4 &= -9.592 \times 10^{-4} \\
                \chi_{r}^2 &= 0.127 \\ 
                \end{aligned} $ & 
                $ \begin{aligned}[c]
                a_1 &= 1.257 \\
                a_2 &= 6.625 \times 10^{-3} \\ 
                a_3 &= 3.227 \times 10^{-2} \\
                a_4 &= -1.698 \times 10^{-3} \\
                \chi_{r}^2 &= 0.122 \\ 
                \end{aligned} $ \\
                \hline
                $ 5 \times 10^{-2} $ & 
                $ \begin{aligned}[c]
                a_1 &= 1.206 \\
                a_2 &= 4.868 \times 10^{-2} \\ 
                a_3 &= 2.484 \times 10^{-2} \\ 
                a_4 &= -1.327 \times 10^{-3} \\ 
                \chi_{r}^2 &= 0.124 \\
                \end{aligned} $ & 
                $ \begin{aligned}[c]
                a_1 &= 1.213 \\
                a_2 &= 3.948 \times 10^{-2} \\ 
                a_3 &= 2.659 \times 10^{-2} \\
                a_4 &= -1.418 \times 10^{-3} \\
                \chi_{r}^2 &= 0.124 \\ 
                \end{aligned} $ & 
                $ \begin{aligned}[c]
                a_1 &= 1.249 \\
                a_2 &= 1.218 \times 10^{-2} \\ 
                a_3 &= 3.127 \times 10^{-2} \\
                a_4 &= -1.647 \times 10^{-3} \\
                \chi_{r}^2 &= 0.122 \\ 
                \end{aligned} $ &
                $ \begin{aligned}[c]
                a_1 &= 1.273 \\
                a_2 &= -2.283 \times 10^{-3} \\ 
                a_3 &= 3.362 \times 10^{-2} \\
                a_4 &= -1.759 \times 10^{-3} \\
                \chi_{r}^2 &= 0.121 \\ 
                \end{aligned} $ \\
                \hline
                \multicolumn{5}{c}{$ \mbox{GR:}
                    a_1 = 1.275,
                    a_2 = -3.584 \times 10^{-3},
                    a_3 = 3.383 \times 10^{-2},
                    a_4 = -1.769 \times 10^{-3};
                    \chi_{r}^2 = 0.121
                    $}\\
                \hline
        \end{tabular} }
        \caption{ { \footnotesize $ I/M^3 $ fit coefficients and corresponding $ \chi^{2} $ scores. The first row and column hold all the presented in the paper values for $ m_{\varphi} $ and $ \lambda $. The entries in each cell present the numerical value of coefficients in \ref{bar_I} and the corresponding $ \chi^{2} $ } }
        \label{Table_uni_coeff_barI}
    \end{table}
    
    \begin{table}
        \makebox[\textwidth][c]{ \begin{tabular}{ c | c | c | c | c }
                \backslashbox{m}{$ \lambda $} & $ 0 $ & $ 1 \times 10^{-1} $ & $ 1 \times 10^{0} $ & $ 1 \times 10^{1} $\\
                \hline
                $ 0 $ & 
                $ \begin{aligned}[c]
                <&L> = 2.162 \times 10^{-2} \\
                <&L_{\infty}> = 5.971 \times 10^{-2} \\ 
                &L_{\infty} = 1.302 \times 10^{-1}\\
                \end{aligned} $ & 
                $ \begin{aligned}[c]
                <&L> = 2.249 \times 10^{-2} \\
                <&L_{\infty}> = 4.922 \times 10^{-2} \\ 
                &L_{\infty} = 1.145 \times 10^{-1}\\
                \end{aligned} $ & 
                $ \begin{aligned}[c]
                <&L> = 2.268 \times 10^{-2} \\
                <&L_{\infty}> = 5.020 \times 10^{-2} \\ 
                &L_{\infty} = 9.745 \times 10^{-2}\\
                \end{aligned} $ &
                $ \begin{aligned}[c]
                <&L> = 2.216 \times 10^{-2} \\
                <&L_{\infty}> = 5.198 \times 10^{-2} \\ 
                &L_{\infty} = 9.286 \times 10^{-2}\\
                \end{aligned} $\\ 
                \hline
                $ 5 \times 10^{-3} $ & 
                $ \begin{aligned}[c]
                <&L> = 2.173 \times 10^{-2} \\
                <&L_{\infty}> = 5.918 \times 10^{-2} \\ 
                &L_{\infty} = 1.259 \times 10^{-1}\\
                \end{aligned} $ & 
                $ \begin{aligned}[c]
                <&L> = 2.227 \times 10^{-2} \\
                <&L_{\infty}> = 4.769 \times 10^{-2} \\ 
                &L_{\infty} = 1.123 \times 10^{-1}\\
                \end{aligned} $ & 
                $ \begin{aligned}[c]
                <&L> = 2.161 \times 10^{-2} \\
                <&L_{\infty}> = 4.229 \times 10^{-2} \\ 
                &L_{\infty} = 9.928 \times 10^{-2}\\
                \end{aligned} $ & 
                $ \begin{aligned}[c]
                <&L> = 2.162 \times 10^{-2} \\
                <&L_{\infty}> = 4.079 \times 10^{-2} \\ 
                &L_{\infty} = 9.418 \times 10^{-2}\\ 
                \end{aligned} $ \\
                \hline
                $ 5 \times 10^{-2} $ & 
                $ \begin{aligned}[c]
                <&L> = 2.081 \times 10^{-2} \\
                <&L_{\infty}> = 4.526 \times 10^{-2} \\ 
                &L_{\infty} = 9.481 \times 10^{-2}\\
                \end{aligned} $ & 
                $ \begin{aligned}[c]
                <&L> = 2.122 \times 10^{-2} \\
                <&L_{\infty}> = 4.453 \times 10^{-2} \\ 
                &L_{\infty} = 9.388 \times 10^{-2}\\
                \end{aligned} $ & 
                $ \begin{aligned}[c]
                <&L> = 2.138 \times 10^{-2} \\
                <&L_{\infty}> = 4.145 \times 10^{-2} \\ 
                &L_{\infty} = 9.377 \times 10^{-2}\\
                \end{aligned} $ &
                $ \begin{aligned}[c]
                <&L> = 2.156 \times 10^{-2} \\
                <&L_{\infty}> = 4.078 \times 10^{-2} \\ 
                &L_{\infty} = 9.406 \times 10^{-2}\\
                \end{aligned} $ \\
                \hline
                \multicolumn{5}{c}{$ \mbox{GR:}
                    <L> = 2.158 \times 10^{-2},
                    <L_{\infty}> = 4.077 \times 10^{-2},
                    L_{\infty} = 9.428 \times 10^{-2}
                    $}\\
                \hline
        \end{tabular} }
        \caption{ { \footnotesize Summary of various averaged norms of $ I/R^3 $ fit residuals $ \left| 1 - \tilde I/\tilde I_{\mbox{fit}}\right| $. The first row and column hold all the presented in the paper values for $ m_{\varphi} $ and $ \lambda $. The entries in each cell present for each fit as follows - $ <L_{1}> $, the average over all EOSs of all the residuals; $ <L_{\infty}> $, the average over all EOSs of the largest relative deviation between the data and fit; $ L_{\infty} $, the largest residual across all EOSs. Only the GR $ <L_{\infty}> $ and $ L_{\infty} $ are included in Fig. \ref{Fig_Uni_bar} as shaded areas.} }
        \label{Table_uni_resd_barI} 
    \end{table}
    
\end{document}